\documentclass[a4paper,11pt]{article}
\pdfoutput=1 

\usepackage{jheppub} 

\usepackage[T1]{fontenc} 

\usepackage{amsthm}

\title{\boldmath Could dark matter be a natural consequence of
a dynamical universe?}

\author[a,b]{Zhi-Wei Wang and }
\author[b]{Samuel L.\ Braunstein}

\affiliation[a]{College of Physics, Jilin University,\\
Changchun, 130012, People's Republic of China}
\affiliation[b]{Computer Science, University of York,\\ York YO10 5GH,
 United Kingdom}

\emailAdd{zhiweiwang.phy@gmail.com}
\emailAdd{sam.braunstein@york.ac.uk}

\abstract{ We construct the {\it gravitating mass\/} of an isolated
composite system on as\-ymp\-tot\-i\-cal\-ly-flat spacetimes within
conventional general relativity and investigate when this quantity is
well defined. For stationary spacetimes, this quantity is known to
exactly equal the {\it physical\/} (ADM) {\it mass}. However, it remains
an open question whether these two masses are equal in the absence of a
timelike Killing vector. This is especially apropos since our universe
has an `origin' and hence no such Killing vector. Further, if these
masses failed to agree then composite systems could behave as if they
had a `dark component,' whose gravitating mass would not equal the
physical mass-energy present. The existence of such an apparent
discrepancy is indeed ubiquitous in galaxies and galaxy clusters, though
currently it is attributed to the presence of dark matter. We conclude
that the theoretical question of the relation between these masses for
dynamical spacetimes is ripe for attention.}

\keywords{Dark matter, 
Classical Theories of Gravity,
Differential and Algebraic Geometry,
Space-Time Symmetries}

\begin{document} 
\maketitle
\flushbottom

\section{Introduction}
\label{sec:Introduction}

Our understanding of general relativity has been largely predicated on
the behavior of stationary spacetimes. Such spacetimes admit a timelike
Killing vector whose algebraic properties allow for significant
simplifications. For example, since 1978 it has been known that the
Komar mass and the physical (ADM) mass are equal for stationary
spacetimes \cite{Beig1978}. Indeed, for such spacetimes the Komar mass
may be interpreted as the mass producing gravitational attraction
assuming an asymptotically Newtonian inverse-square law. Thus, for more
than forty years, it has been taken as fact in general relativistic
systems, that the gravitating mass and the physical mass are identical.
But does this result really hold for dynamical spacetimes? Indeed, in
the absence of a timelike Killing vector the Komar mass is not even
considered a legitimate well-defined measure of mass for dynamical
spacetimes \cite{wald1984}. However, the untested assumption of the
equality between the gravitating and physical masses for such generally
dynamical composite systems lies at the heart of the dark matter
paradigm.


Here we begin to address this foundational issue by formulating the
gravitating mass for arbitrary systems in generally dynamical
spacetimes. We show that this quantity is in fact well-defined for a
broad class of asymptotically-flat dynamical spacetimes (i.e., where no
Killing vector exists). This analysis therefore allows us to precisely
formulate for the first time the question of whether conventional
general relativity really does predict the equivalence between
gravitating mass and physical mass-energy for dynamical composite
systems.

\section{Gravitating mass for dynamical spacetimes}
\label{sec:Gravitating}

Consider an arbitrary smooth foliation of spacetime into a family of
spacelike hypersurfaces. Such a foliation singles out a unique unit
timelike covector field $\hat T_\mu$ whose kernel is the tangent space
on each hypersurface \cite{Frobenius1877}. These covectors are
dual to a vector field $\hat T^\mu$ which may be interpreted as
4-velocities of a uniquely specified family of canonical observers with
regard to that family of hypersurfaces. We will use the 4-acceleration
\begin{equation}
a^\mu={\hat T^\mu}_{~~;\nu} \hat T^\nu,
\end{equation}
of these canonical observers to
`measure' the gravitating mass-energy within a closed surface, $S$, on
any specific hypersurface $\Sigma$ within the family, by
\begin{equation}
M_{\text{Grav}}=
\frac{1}{4\pi}\!
\int_S {\cal N}a^\mu  \hat N_\mu\, dA
=\frac{1}{4\pi}\!
\int_S {\cal N}\,\hat T^\mu_{~~;\nu} \hat T^\nu  \hat N_\mu\, dA.
\label{grav_mass}
\end{equation}
In the stationary setting,
Eq.~(\ref{grav_mass}) is called the Komar mass \cite{Komar1959}.
Here, the inclusion of the lapse function, ${\cal N}$, is to guarantee
that energies/forces are quantified with regard to the values that would
be seen at spatial infinity. In addition, here $dA$ is the area element
on $S$ and $\hat N^\mu$ is the outward pointing unit normal to $S$
tangent to the hypersurface $\Sigma$. For example, for the Schwarzschild
metric of a black hole of mass $M$, this expression
Eq.~(\ref{grav_mass}) yields
\begin{equation}
M_{\text{Grav}}=M,
\end{equation}
for any $S$ containing the black hole.

\begin{figure}[tbp]
\centering
\includegraphics[width=0.6\textwidth]{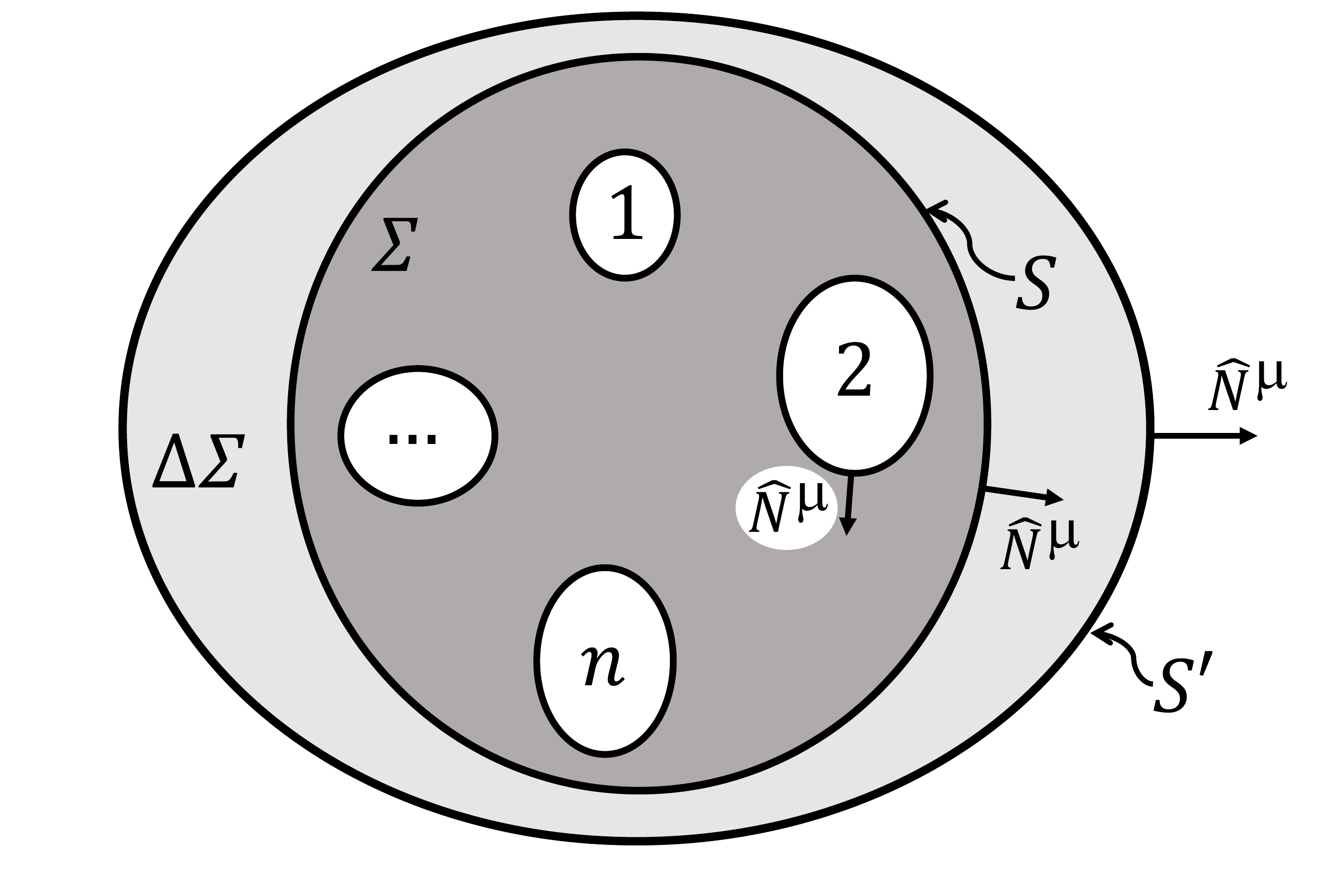}
\caption{For $M_{\text{Grav}}$ to be well-defined, it must be
independent of the choice of boundary $S$. Here the portion of hypersurface 
$\Delta\Sigma$ is bounded by $S$ and $S'$.}
\label{fig1}
\end{figure}

Of course, for $M_{\text{Grav}}$ to be physically meaningful, 
we would require that the expression given in
Eq.~(\ref{grav_mass}) be independent of the choice of surface $S$.
Let us now explore the implications of this constraint.
\begin{eqnarray}
&&\frac{1}{4\pi}\!
\int_S {\cal N}\,\hat T^\mu_{~~;\nu} \hat T^\nu  \hat N_\mu\, dA 
\nonumber \\
&=& \frac{1}{4\pi}\!
\int_S ({\cal N}\,\hat T^{[\mu})^{;\nu]} \hat N_\mu \hat T_\nu \, dA
\nonumber \\
&=& \frac{1}{4\pi}\!
\int_{S'} {\cal N}\,\hat T^\mu_{~~;\nu} \hat T^\nu  \hat N_\mu\, dA
- \frac{1}{4\pi}\!
\int_{\Delta\Sigma} \!\!({\cal N}\,\hat T^{[\mu})^{;\nu]} _{\;\;\;\; ;\mu}
\hat T_\nu \, dV, \nonumber \\
\label{two boundary}
\end{eqnarray}
where $S'$ is a new closed surface on the hypersurface $\Sigma$,
$\Delta\Sigma$ is a portion of the hypersurface bounded at either end by
$S$ and $S'$ (see Fig.~\ref{fig1}), and $dV$ is the volume element on
$\Sigma$. We use Lemma 1 for the first step and Stokes' theorem for the
final step.

\vskip 0.1in
\noindent
{\bf Lemma 1:} For a spacelike hypersurface $\Sigma$ with tangent
vector $\hat N^\mu$ and unit normal $\hat T^\mu$ we have
\begin{equation}
{\cal N}\,\hat T^\mu_{~~;\nu} \hat T^\nu  \hat N_\mu
= ({\cal N}\,\hat T^{[\mu})^{;\nu]} \hat N_\mu \hat T_\nu
\end{equation}
where ${\cal N}$ is the lapse function.

\vskip 0.1in
\noindent
{\bf Proof:} The foliation structure guarantees that our covector takes
the form $\hat T_\mu\propto t_{,\mu}$, where $t$ is a function on the
spacetime that labels individual members of the foliation. Thus,
$t$ corresponds to a `time' coordinate \cite{Natario}. 
Normalizing $\hat T_\mu$, and orienting it to be future directed, then gives
$\hat T_\mu=-{\cal N} t_{,\mu}=-{\cal N} t_{;\mu}$ and hence
\begin{eqnarray}
\hat T_{\mu ; \nu} &=& -({\cal N} t_{;\mu})_{; \nu}
= -{\cal N} _{; \nu} t_{;\mu} -
{\cal N} t_{;\mu \nu} \nonumber \\
&=& {\cal N} _{; \nu} \frac{1}{{\cal N}}
\hat T_\mu -{\cal N} t_{;\nu \mu} \nonumber \\
&=& \frac{1}{{\cal N}}\, {\cal N} _{; \nu} 
\hat T_\mu +{\cal N} ( \frac{1}{{\cal N}}\, \hat T_\nu )_{; \mu}
 \nonumber \\
&=& \frac{1}{{\cal N}}\, {\cal N} _{; \nu} 
\hat T_\mu -\frac{1}{{\cal N}}\, {\cal N}_{; \mu}
\hat T_\nu + \hat T_{\nu ; \mu}  ,
\label{T_mu_nu}
\end{eqnarray}
where since $t$ is a scalar we have used $t_{;\mu \nu} = t_{; \nu \mu}$
in the first line. Note that 
$\hat T_{[\mu,\nu]}=\hat T_{[\mu}{\cal N}_{,\nu]}/{\cal N}$
follows directly from Eq.~(\ref{T_mu_nu}); a form
which is consistent with the existence of our
foliation \cite{Frobenius1877,Natario,poisson2004}.

From Eq.~(\ref{T_mu_nu}), the 4-acceleration $a_\mu$ may be written as
\begin{eqnarray}
a_\mu &\equiv& \hat T_{\mu ; \nu} \hat T^\nu  \nonumber \\
&=& \frac{1}{{\cal N}}\, {\cal N} _{; \nu} \hat T^\nu
\hat T_\mu -\frac{1}{{\cal N}}\, {\cal N}_{; \mu} \hat T^\nu
\hat T_\nu + \hat T_{\nu ; \mu} \hat T^\nu \nonumber \\
&=& \frac{1}{{\cal N}}\, {\cal N} _{; \nu} \hat T^\nu
\hat T_\mu +\frac{1}{{\cal N}}\, {\cal N}_{; \mu} 
= \frac{1}{\cal N}\, {\cal N}_{ ; \nu} {h^\nu}_\mu   ,
\label{2ndlast} 
\end{eqnarray}
where we have used $\hat T_{\nu ; \mu} \hat T^\nu=0$ in the second line
and $h_{\mu\nu}\equiv g_{\mu\nu} + \hat T_\mu \hat T_\nu$ in the
last step.

Finally, from Eq.~(\ref{2ndlast}),
$({\cal N} \hat T_{[\mu})_{; \nu]} \hat T^\nu  \hat N^\mu $
may be simplified as
\begin{eqnarray}
&& ({\cal N} \hat T_{[\mu})_{; \nu]}
\hat T^\nu  \hat N^\mu \nonumber \\
&=&\frac{1}{2} ({\cal N} \hat T_\mu)_{ ; \nu}
\hat T^\nu  \hat N^\mu - \frac{1}{2}({\cal N}
\hat T_\nu)_{ ; \mu} \hat T^\nu  \hat N^\mu \nonumber \\
&=&\frac{1}{2} {\cal N}_{ ; \nu} \hat T_\mu
\hat T^\nu  \hat N^\mu + \frac{1}{2} {\cal N} \hat T_{\mu ; \nu} \hat T^\nu
 \hat N^\mu - \frac{1}{2} {\cal N}_{ ; \mu} \hat T_\nu \hat T^\nu  \hat N^\mu
- \frac{1}{2} {\cal N} \hat T_{\nu ; \mu}
\hat T^\nu  \hat N^\mu  \nonumber \\
&=&\frac{1}{2} {\cal N} \hat T_{\mu ; \nu} \hat T^\nu
 \hat N^\mu - \frac{1}{2} {\cal N}_{ ; \mu} \hat T_\nu \hat T^\nu 
\hat N^\mu \nonumber \\
&=& {\cal N} \hat T_{\mu ; \nu} \hat T^\nu
 \hat N^\mu  ,
 \label{Liexi=0}
\end{eqnarray}
where we have used $\hat T_{\nu ; \mu} \hat T^\nu=0$ in the
third line, and Eq.~(\ref{2ndlast}) in the second-last line.

This completes the proof of Lemma 1.
\qed
\vskip 0.1in

From Eq.~(\ref{two boundary}) we know that if we want $M_{\text{Grav}}$
to be independent of the choice of $S$, we require
that the volume integral between these two boundaries must vanish, i.e.,
\begin{eqnarray}
 \frac{1}{4\pi}\!
\int_{\Delta\Sigma}\!\! ({\cal N}\,\hat T^{[\mu})^{;\nu]} _{\;\;\;\; ;\mu}
\hat T_\nu \, dV =0 .
\label{volume}
\end{eqnarray}
To simplify this condition, we use:

\vskip 0.1in
\noindent
{\bf Lemma 2:} For a vector field $\xi^\mu$ (although 
$\xi^\mu={\cal N} \hat T^\mu$ is used for this paper), we have
\begin{equation}
\xi_{[\beta;\alpha];\nu}
+ \frac{1}{2} ( \mathfrak{L }_\xi g_{\nu \alpha} )_{;\beta}
- \frac{1}{2} (\mathfrak{L }_\xi g_{\beta \nu} )_{;\alpha}
= R_{\mu\nu\alpha\beta} \xi^\mu ,
\label{L3}
\end{equation}
where $\mathfrak{L }_\xi$ is the Lie derivative along $\xi^\mu$.

\vskip 0.1in
\noindent
{\bf Proof:}
Since $\xi_{\nu;\alpha\beta} - \xi_{\nu;\beta \alpha}
= R_{\mu \nu \alpha \beta} \xi^\mu$, the
left-hand-side of Eq.~(\ref{L3}) may be simplified as
\begin{eqnarray}
    &=&  \xi_{[\beta;\alpha];\nu}
+ \xi_{(\nu;\alpha);\beta} - \xi_{(\beta;\nu );\alpha} \nonumber  \\
    &=& \frac{1}{2} \Bigl( \xi_{\beta ;\alpha \nu }
- \xi_{\alpha ;\beta \nu } + \xi_{\nu ; \alpha \beta}
+ \xi_{\alpha ;\nu \beta} - \xi_{\beta ;\nu \alpha}
 - \xi_{\nu ; \beta \alpha} \Bigr)  \nonumber  \\
    &=&  \frac{1}{2}\Bigl[ (\xi_{\beta ;\alpha \nu } 
- \xi_{\beta ;\nu \alpha})\! +\!   (\xi_{\nu ; \alpha \beta}
 - \xi_{\nu ; \beta \alpha}) \!+\!  (\xi_{\alpha;\nu \beta}
- \xi_{\alpha ;\beta \nu }) \Bigr] \nonumber  \\
    &=& \frac{1}{2} \Bigl( R_{\mu \beta \alpha \nu} \xi^\mu
+ R_{\mu \nu \alpha \beta} \xi^\mu  + R_{\mu \alpha \nu \beta} \xi^\mu \Bigr)  
    \nonumber  \\
    &=&  \frac{1}{2} ( R_{\mu \nu \alpha \beta} \xi^\mu
- R_{\mu \nu \beta \alpha} \xi^\mu ) 
    = R_{\mu \nu \alpha \beta} \xi^\mu ,
\label{Rmn1}
\end{eqnarray}
where the Bianchi identity,
$R_{\mu \alpha \nu \beta} + R_{\mu \nu \beta \alpha}
+ R_{\mu \beta \alpha \nu} = 0$, is used in the fourth line to obtain
the fifth line.

This completes the proof of Lemma 2.
\qed
\vskip 0.1in 

Contracting $\nu$ with $\beta$ in Eq.~(\ref{L3}) and replacing $\alpha$
by $\nu$ yields
\begin{eqnarray}
    {\xi_{[\mu;\nu]}}^{;\mu} = R_{\mu \nu} \xi^\mu
-\frac{1}{2} ( \mathfrak{L }_\xi g_{\mu \nu} )^{;\mu}
+ \frac{1}{2} ( \mathfrak{L }_\xi g_{\alpha\beta} )_{;\nu} g^{\alpha \beta}. 
\label{Rmn5}
\end{eqnarray}

\noindent
{\bf Corollary:} When ${\cal N}\hat T^\mu$ is a Killing vector (and by
construction, orthogonal to the hypersurface containing the boundary
$S$) then $M_{\text{Grav}}$ is independent of variations of $S$ through
any matter-free region of space.

\vskip 0.1in
This follows straightforwardly by taking $\xi^\mu={\cal N}\hat T^\mu$,
since from Eq.~(\ref{Rmn5}) the conditions of the corollary immediately
imply that the condition in Eq.~(\ref{volume}) is exactly satisfied
(since all Lie derivatives, $\mathfrak{L }_\xi$, vanish when $\xi^\mu$
is Killing). Indeed, this is the standard text-book result
for stationary spacetimes \cite{wald1984}. To go beyond this for
dynamical spacetimes we make use of:

\vskip 0.1in
\noindent
{\bf Lemma 3:} For $\xi^\mu={\cal N}\hat T^\mu$ we have
\begin{equation}
 \frac{1}{2}\bigl[ ( \mathfrak{L }_\xi g_{\alpha\beta} )_{;\nu}
g^{\alpha \beta}
-( \mathfrak{L }_\xi g_{\mu \nu} )^{;\mu}\bigr] \hat T^\nu
={\cal N}(K_{,\nu}\hat T^\nu +K_{\mu\nu}K^{\mu\nu}),
\end{equation}
where $K_{\mu\nu}={h_\mu}^\alpha {h_\nu}^\beta \,\hat T_{(\alpha;\beta)}$
is the extrinsic curvature and
$K=g^{\mu\nu}K_{\mu\nu}=h^{\mu\nu}K_{\mu\nu}=\hat T^\mu_{~~;\mu}$
is its trace.

\vskip 0.1in
\noindent
{\bf Proof:} Taking $\xi^\mu={\cal N}\hat T^\mu$ we find the first term
of the Lemma may be rewritten as
\begin{eqnarray}
&&\frac{1}{2} ( \mathfrak{L }_\xi g_{\alpha\beta} )_{;\nu} g^{\alpha \beta}
=({\cal N} \hat T_{(\alpha})_{;\beta);\nu} g^{\alpha\beta}\nonumber \\
&=&\bigl({\cal N}_{;(\alpha}\hat T_{\beta)}g^{\alpha \beta}
+{\cal N}\hat T_{(\alpha;\beta)}g^{\alpha \beta} \bigr)_{;\nu} 
\nonumber \\
&=& \bigl({\cal N}_{;\alpha}\hat T^\alpha
+{\cal N}(-\hat T^\alpha \hat T^\beta+h^{\alpha\beta}) 
\hat T_{(\alpha;\beta)} \bigr)_{;\nu} \nonumber \\
&=& \bigl({\cal N}_{;\mu}\hat T^\mu +{\cal N} K \bigr)_{;\nu}.
\label{extra1}
\end{eqnarray}

In a similar manner we find
\begin{eqnarray}
&&-\frac{1}{2} ( \mathfrak{L }_\xi g_{\mu \nu} )^{;\mu} = -( {\cal N} \hat T_{(\mu })_{;\nu)}{}^{ ;\mu} \nonumber \\
&=& \bigl(-{\cal N} \hat T_{(\mu;\nu)} -{\cal N}_{;(\mu} \hat T_{\nu)} \bigr)^{;\mu}
\nonumber \\
&=& \bigl( -{\cal N} \hat T_{(\alpha;\beta)} (-\hat T^\alpha \hat T_\mu + {h^\alpha}_\mu)(-\hat T^\beta \hat T_\nu + {h^\beta}_\nu) -{\cal N}_{;(\mu} \hat T_{\nu)} \bigr)^{;\mu}
\nonumber \\
&=&\bigl( {\cal N}\hat T_{(\mu}a_{\nu)}-{\cal N}_{;(\mu} \hat T_{\nu)}
-{\cal N}K_{\mu\nu}\bigr)^{;\mu} \nonumber \\
&=& \bigl( {\cal N}\hat T_{(\mu}a_{\nu)}
 - {\cal N} a_{(\mu} \hat T_{\nu)}
 + {\cal N}_{; \lambda} \hat T^\lambda \hat T_{(\mu} \hat T_{\nu)}
 -{\cal N}K_{\mu\nu}\bigr)^{;\mu} \nonumber \\
&=& \bigl( {\cal N}_{; \lambda} \hat T^\lambda \hat T_\mu \hat T_\nu
 -{\cal N}K_{\mu\nu}\bigr)^{;\mu},
\label{extra22}
\end{eqnarray}
where we use $a_\mu \hat T^\mu=0$ and $a_\mu h^\mu_{~~\nu}=a_\nu$ to
obtain the fourth line, and ${\cal N}_{; \mu}= {\cal N} a_\mu - {\cal
N}_{;\nu} \hat T^\nu \hat T_\mu $ from Eq.~(\ref{2ndlast}) to obtain the
fifth line.

Now combining Eqs.~(\ref{extra1}) and (\ref{extra22}), we have 
\begin{eqnarray}
&&\frac{1}{2}\bigl[ ( \mathfrak{L }_\xi g_{\alpha\beta} )_{;\nu}
g^{\alpha \beta}
-( \mathfrak{L }_\xi g_{\mu \nu} )^{;\mu}\bigr] \hat T^\nu
\nonumber \\
&=&  \bigl({\cal N}_{;\mu}\hat T^\mu +{\cal N} K \bigr)_{;\nu} \hat T^\nu
  + \bigl( {\cal N}_{; \lambda} \hat T^\lambda \hat T_\mu \hat T_\nu
 -{\cal N}K_{\mu\nu}\bigr)^{;\mu} \hat T^\nu \nonumber \\
&=&  \bigl({\cal N}_{;\mu}\hat T^\mu \bigr)_{;\nu} \hat T^\nu
 + \bigl({\cal N} K \bigr)_{;\nu} \hat T^\nu
 + \bigl( {\cal N}_{; \lambda} \hat T^\lambda \bigl)^{;\mu}
 \hat T^\nu \hat T_\mu \hat T_\nu + {\cal N}_{; \lambda} \hat T^\lambda \bigl( \hat T_\mu \hat T_\nu
 \bigl)^{;\mu} \hat T^\nu - \bigl( {\cal N}K_{\mu\nu}\bigr)^{;\mu}
 \hat T^\nu  \nonumber \\
&=& \bigl({\cal N} K \bigr)_{;\nu} \hat T^\nu
 + {\cal N}_{; \lambda} \hat T^\lambda \bigl( \hat T_\mu \bigl)^{;\mu}
 \hat T^\nu \hat T_\nu - \bigl( {\cal N}K_{\mu\nu}\bigr)^{;\mu} \hat T^\nu
\nonumber \\
&=& \bigl({\cal N} K \bigr)_{;\nu} \hat T^\nu - {\cal N}_{; \lambda}
 \hat T^\lambda K - \bigl( {\cal N}K_{\mu\nu}\bigr)^{;\mu}
 \hat T^\nu  \nonumber \\
&=& {\cal N} K_{;\nu} \hat T^\nu -{\cal N}K_{\mu\nu}^{~~~;\mu} \hat T^\nu 
\nonumber \\
&=& {\cal N}K_{,\nu} \hat T^\nu
-{\cal N}\bigl(K_{\mu\nu} \hat T^\nu \bigr)^{;\mu}
+{\cal N}K_{\mu\nu} \hat T^{\nu;\mu}
\nonumber \\
&=& {\cal N} K_{,\nu}\hat T^\nu +{\cal N}K_{\mu\nu}K^{\mu\nu} \! , 
\label{extra23}
\end{eqnarray}
where we use $K=\hat T^\mu_{~~;\mu}$ and $K_{\mu\nu}\hat T^\nu=0$.

This completes the proof of Lemma 3.
\qed
\vskip 0.1in 

We may now state our main result:

\vskip 0.1in
\noindent
{\bf Theorem:} On asymptotically matter-free, $T_{\mu\nu}=o(1/r^3)$,
asymptotically-flat {\it dynamical\/} spacetimes
\begin{equation}
M_{\text{Grav}}\equiv\lim_{S\rightarrow i^0}
\frac{1}{4\pi}\!
\int_S {\cal N}\,\hat T^\mu_{~~;\nu} \hat T^\nu  \hat N_\mu\, dA,
\label{thm}
\end{equation}
provides a well-defined measure of gravitating mass on the hypersurface
within the closed surface $S$ as it is taken to spatial infinity, $i^0$,
provided $K_{,\mu} \hat T^\mu=o(1/r^3)$. (Here, $r$ is the asymptotic
radial coordinate and $g_{\mu\nu}=\eta_{\mu\nu}+O(1/r)$,
$g_{\mu\nu,\sigma}=O(1/r^2)$, etc. Recall the little-o notation
$f(r)=o(1/r^n)$ means $\lim_{r\rightarrow \infty} r^n f(r)=0$.)

\vskip 0.1in
\noindent
{\bf Proof:}
In order to prove our theorem, we must show that the limit is
independent of the choice of boundary, $S$, as it limits to spatial
infinity. In order to do this, we need only show that volume term in
Eq.~(\ref{two boundary}) vanishes asymptotically. Using the integrand
expressed as Eq.~(\ref{Rmn5}) and Lemma 3, we may write this volume term
as
\begin{eqnarray}
 && \frac{1}{4\pi}\!
\int_{\Delta\Sigma}\!\! ({\cal N}\,\hat T^{[\mu})^{;\nu]} _{\;\;\;\; ;\mu}
\hat T_\nu \, dV \label{volume4} \\ 
&=& \frac{1}{4\pi}\! \int_{\Delta\Sigma}\!\! {\cal N}
 \bigl( R_{\mu \nu} \hat T^\mu \hat T^\nu
 + K_{,\mu}\hat T^\mu + K_{\mu\nu}K^{\mu\nu} \bigr)\, dV.
\nonumber
\end{eqnarray}

For an asymptotically matter-free spacetime with $T_{\mu \nu}=o(1/r^3)$,
we see that $R_{\mu \nu}=o(1/r^3)$ so this term will not contribute
asymptotically. Further, $K_{\mu\nu}=O(1/r^2)$ so
$K_{\mu\nu}K^{\mu\nu}=O(1/r^4)$ and this term too will make no
asymptotic contribution. This only leaves the term $K_{,\mu}\hat T^\mu$.
Ordinarily the asymptotically-flat constraint would imply that
$K_{,\mu}=O(1/r^3)$ and so an asymptotic contribution would be
possible from this remaining term. However, if we modestly strengthen
this constraint to be $K_{,\mu}\hat T^\mu=o(1/r^3)$, the
integral in Eq.~(\ref{volume4}) now vanishes asymptotically and
consequently, Stokes' theorem and Eq.~(\ref{two boundary}) guarantee
that $M_{\text{Grav}}$ is asymptotically independent of the choice of
boundary as it is taken to spatial infinity.

This completes the proof for our theorem.
\qed

\vskip 0.1in

We emphasize that at no stage has the existence of a Killing
vector of any kind been assumed to obtain this result. Further, our
asymptotic conditions are only required to hold at spatial infinity,
$i^0$. In particular, we place no constraints on the behavior at null
infinity. Thus, our dynamical composite system may potentially be
extremely violent on the interior of the spacelike hypersurface,
$\Sigma$, for example, generating significant gravitational radiation.

Thus, we have shown that the gravitating mass is well-defined even for a
broad class of asymptotically-flat dynamical spacetimes. Note that the
gravitating mass, $M_{\text{Grav}}$, is evaluated on a closed surface
$S$ at spatial infinity $i^0$ and is generally only well defined when
evaluated there (just as is the case with the ADM mass). As already
noted, the gravitating and ADM mass are known to be equal for stationary
spacetimes, though their relationship is far less clear in the dynamical
setting.

The mathematical form of $M_{\text{Grav}}$ in Eq.~(\ref{thm}) is
identical to that of the Bondi mass \cite{Tamburino66} except that the
latter is evaluated on a null hypersurface at null infinity. When the
metric under consideration may be reduced to Bondi form at null infinity
the Bondi mass gives a measure of the energy contained in null (e.g.,
gravitational) radiation. Further, assuming such an asymptotic-Bondi
form and in the absence of gravitational radiation, the infinite past
limit of the Bondi mass is known to equal the ADM mass \cite{Zhang06}.
Thus, one might be tempted to take the position that it is likely that
the gravitating mass evaluated on a spacelike hypersurface and the ADM
mass are identical even in the dynamical setting. However, i) the
gravitating mass may be well defined even when the Bondi mass is not
(since the assumed form of the asymptotic metric at null infinity may
not hold) and ii) the relation between the Bondi mass and ADM mass is
known to break down for dynamical spacetimes \cite{Zhang06}. If
anything, one might instead expect a `mass anomaly' whereby for
dynamical spacetimes the gravitating mass and ADM mass generally do not
agree. We now explore potential implications for dark matter were such a
mass anomaly to exist.

\section{Possible implications for dark matter}
\label{sec:implications}

The orbital motion of stars within their host galaxy and even the motion
of galaxies within galaxy clusters cannot be understood without some
extra source of gravitational attraction other than that due to the
visible matter. This extra attraction is widely accepted to be caused by
`dark matter,' with the current prevailing model being that of cold dark
matter \cite{Bertone2018}. 

However, the failure to directly detect the particles responsible for
dark matter has led to alternative theories to explain these unexpected
gravitational phenomena. The most successful of these is known as
modified Newtonian dynamics (MOND) and its variations \cite{Milgrom1983a}.
Such modifications to Newtonian dynamics are
challenged by scenarios involving collisions between clusters of
galaxies, such as in the Bullet Cluster \cite{Markevitch04}. More recent
evidence suggests that any modifications to general relativity capable
of explaining dark matter would be inconsistent with large-scale
cosmological observations \cite{Pardo20}.

Here we suggest a third option is worth investigating: In particular we
ask, within conventional general relativity, whether the signals for
dark matter might be understood in terms of the normative behavior
of dynamical composite systems. For dynamical spacetimes there is
currently no proof that the gravitating mass of a composite system
equals the amount of physical mass there. Although these two masses are
known to be equal for stationary spacetimes, this result does not apply
to our own expanding universe, where stationarity is ruled out. A
breakdown of equality between these masses would signal a `mass anomaly'
which may partly explain the observed discrepant motion of stars and
galaxies.

Unlike the physical (ADM) mass which is based on the Hamiltonian
formalism, the concept of gravitating mass has previously been limited
to stationary spacetimes where it reduces to the Komar
mass \cite{wald1984}. The major obstacle to extending this work has been
that the Komar mass requires a timelike Killing vector which is by
definition absent in dynamical spacetimes. In the stationary setting,
the timelike Killing vector field yields the 4-velocities for a family
of `static' observers, which may be used to determine the gravitating
mass on hypersurfaces orthogonal to this Killing field \cite{wald1984}.
Here we invert this reasoning and determine the gravitating mass
measured by canonical observers moving orthogonally to a family of
spacelike hypersurfaces. We find that even in the absence of a Killing
field, this approach leads to a well-defined measure of gravitating mass
even for a broad class of dynamical asymptotically-flat spacetimes.
Thus, our work is the first to generalize this concept to dynamical
spacetimes. Further, at least on the scale of galaxies, galaxy clusters
and even super clusters, where the signature of dark matter is
unmistakable, the universe may be well approximated as
flat \cite{Dawson13}. The remaining open question is whether or not, for
such dynamical spacetimes, the gravitating and physical masses are in
agreement.




\end{document}